# SignalNet: A Low Resolution Sinusoid Decomposition and Estimation Network

Ryan M. Dreifuerst, Student Member, IEEE, Robert W. Heath Jr. Fellow, IEEE,

*Abstract*—The detection and estimation of sinusoids is a fundamental signal processing task for many applications related to sensing and communications. While algorithms have been proposed for this setting, quantization is a critical, but often ignored modeling effect. In wireless communications, estimation with low resolution data converters is relevant for reduced power consumption in wideband receivers. Similarly, low resolution sampling in imaging and spectrum sensing allows for efficient data collection. In this work, we propose SignalNet, a neural network architecture that detects the number of sinusoids and estimates their parameters from quantized in-phase and quadrature samples. We incorporate signal reconstruction internally as domain knowledge within the network to enhance learning and surpass traditional algorithms in mean squared error and Chamfer error. We introduce a worst-case learning threshold for comparing the results of our network relative to the underlying data distributions. This threshold provides insight into why neural networks tend to outperform traditional methods and into the learned relationships between the input and output distributions. In simulation, we find that our algorithm is always able to surpass the threshold for three-bit data but often cannot exceed the threshold for one-bit data. We use the learning threshold to explain, in the one-bit case, how our estimators learn to minimize the distributional loss, rather than learn features from the data.

*Index Terms*—Sinusoid decomposition, estimation, deep-learning, low-resolution, detection

## I. Introduction

Detecting the number of sinusoids and estimating the amplitude, frequency, and phase is a common problem in signal processing [1]–[4]. In these settings, information is contained in the sinusoidal parameters of the data, i.e. in the Doppler shift from a radar signal or the angles of arrival and departure in the spatial domain. As bandwidth and array size continue to increase, e.g. in mmWave communications or wideband automotive radar, traditional systems require analog-to-digital converters (ADCs) with high resolution and rates approaching 100 Gbps or more. High resolution data converters become a limiting factor for low power consumption [5]. One solution to reduce power consumption is to reduce the resolution in ADCs, transferring the difficulty of the problem from the replaced power ADCs to practical digital signal processing algorithms [5]–[8].

Ryan M. Dreifuerst is with The University of Texas, Austin, TX 78712 USA (ryandry1st@utexas.edu). Robert W. Heath Jr. is with North Carolina State University, Raleigh, NC 27695 (rwheathjr@ncsu.edu). This work is based upon work supported in part by Samsung Research America and the National Science Foundation under Grant No. ECCS-1711702.

In principle, low resolution techniques improve efficiency–computationally, economically, or in power consumption–at the cost of introducing quantization error. With high resolution sampling the quantization error is often modeled as an additive noise term [6], [9]. This is motivated to linearize the quantization, such as Bussgang Decomposition [10], [11]. Ultimately, the linearization has more error for 1-3 bit quantizers, which are the most desirable from an efficiency perspective. To overcome this limitation, we investigate neural network techniques for low resolution signal processing because of their ability to learn analytical models directly from data. The use cases for such algorithms are broad, with varying compute constraints (server, cloud, deployed), but our algorithm is primarily intended for edge computer signal processing, so we also consider important characteristics such as memory, training sample size, and execution time in our investigation. From our past work, we found convolutional neural networks were sufficiently powerful and fast for near real-time estimation of a single sinusoid [12]. As a result, our networks rely heavily on convolutional layers for parameter sharing, dimensionality reduction, and information parsing.

There is, to the best of our knowledge, little prior work that has focused directly on the joint task of detection and estimation of sinusoids from quantized data. In contrast, the field of unquantized detection and estimation of sinusoids has a rich history [13]–[17]. These techniques tend to divide the problem into separate, but related, tasks for detection and estimation. Detection is usually performed using model order estimators such as Minimum Description Length (MDL) [15], [18] or Akaike Information Criterion (AIC) [14], [19]. Then, given a prediction of the number of sinusoids, the estimation is typically solved using non-parametric or parametric methods [17], [20], [21]. Alternatively, the problem can be approached using compressive sensing [22] techniques. In this paradigm, the recovered signal vector is the frequency representation of the multiple sinusoids, enabling sparse reconstruction techniques. Notable algorithms in this domain are based on the message passing algoritm [23], [24]. These methods are restricted to on-grid measurements, however, which affects the accuracy for limited samples. More recently, an additional class of algorithms based on neural networks have also been tasked with the same problem [25]. This has led to new state of the art results for the detection and estimation of sinusoid frequencies [25]. Neural networks have the advantage of learning directly from training data, so no assumptions regarding the SNR or model order are necessary. Instead, the model learns assumptions and distributions from the training data, possibly leading to inadvertent bias. While inductive bias is necessary for successful learning [26], it can have adverse effects when data distribution changes. Investigation of the understand-ability of neural networks is an active area of research [27].

More prior work has focused on the estimation of a single quantized sinusoid in noise [28]–[31]. The Cramér Rao Bound was derived in [28]. Algorithms using correlation [28], dithering [30], and time-varying thresholding [31] have been proposed for quantized sinusoid estimation. Dithering and time-varying thresholding are related by effectively adjusting the sampling comparator by either a known (time-varying thresholds) or unknown factor (dithering). These methods rely on knowledge of the amplitude of the signal, or the ability to estimate by iterative grid search. Unlike the unquantized case, non-parametric methods, such as the Periodogram, are often the best estimators with extremely low-resolution sinusoid estimation [12], [30]. All of these techniques (non-parametric estimators [21], correlation-based estimators [28], dithering [30] and time-varying thresholds [31]) do not make use of the quantization noise correlation [11]. This correlation is not easily modeled in accurate and tractable forms. Instead, our neural network attempts to learn the underlying relationships directly from data, bypassing the need for explicit models. We also contend with new concepts such as the minimum recognizable separation between signals, which strongly limits the performance of the previously mentioned estimators.

Our paper is a rigorous investigation into deep learning for sinusoidal parameter estimation from low resolution sampling. We propose and analyze a state of the art architecture, SignalNet, that relies on internal reconstruction to successfully estimate multi-sinusoidal representations. Traditional algorithms do not account for quantization effects and instead are applied with Bussgang Decomposition to linearize the nonlinear distortion. This is the defacto approach for other low resolution problems like channel estimation [8], [11]. While it may not be surprising that neural networks are well-suited to handle nonlinear systems, we show that the use of domain knowledge within SignalNet outperforms simple neural networks. Finally, we analyze the proposed network with both statistical measures through a new learning threshold and out-of-distribution data. These tests show that the SignalNet architecture is able to learn meaningful features that generalize well and ensure the neural model is capable of handling crucial signal processing tasks across many domains. Our algorithm follows traditional approaches by dividing the task into two separate networks, which we identify as the detection module and the estimation module. The focal point of our algorithm is the estimation module, which sequentially estimates sinusoid parameters. Within the estimation module, we configure the network to reconstruct the input sequences and feed the error back into the estimation module, thereby explicitly defining the relationship between the output estimates and input sequences. This method instills domain knowledge within the network and helps the model directly capture the input-output relationship for the sinusoid detection and estimation problem. We then benchmark our algorithm against classical methods for each module, as well as define and derive learning thresholds for the problem based on estimation. In simulation, we show that both of our modules outperform traditional algorithms for quantized data. In spite of this, our detection module and amplitude estimation algorithms are not able to surpass the learning threshold for one-bit resolution data. Our contributions can be summarized as follows:

- We propose a novel deep learning algorithm for sinusoid parameter estimation. Our algorithm uses reconstruction as an internal mechanism for learning the relationship between sinusoidal parameters and input data. We show that this method is able to estimate sinusoids significantly more accurately than traditional algorithms for quantized data. We further show that our network generalizes nearly ideally to out-of-distribution datasets.

- We extend our sinusoid parameter estimator to include multiple sinusoid estimation and detection. We find that the overall network, SignalNet, is able to accurately detect the number of sinusoidal signals and their associated parameters from limited observations of quantized data. The proposed algorithm achieves state of the art results and universally improves upon the benchmark algorithms.

- We define a new benchmark for comparing neural network algorithms and traditional algorithms based on the loss function and randomness of the output distribution. The resulting threshold defines the worst-case error expected for non-adversarial data, and provides insight into why neural networks tend to outperform other algorithms within signal processing–even for nearly random data. Applying this threshold to our simulations shows that our one-bit sinusoid detection algorithm and one-bit amplitude estimates are not able to learn meaningful input-output relationships. All other networks surpass the learning threshold and traditional algorithms, suggesting our algorithms are learning substantial information from the data.

Notation: $\mathbf{A}$ is a matrix, $\mathbf{a}$ and $\{a[i]\}_{i=1}^{N}$ are column vectors and $a, A$ denote scalars. $\mathbf{A}^{T}$, $\bar{\mathbf{A}}$ and $\mathbf{A}^{*}$ represent the transpose, conjugate, and conjugate transpose of $\mathbf{A}$. The real and imaginary parts of $\mathbf{A}$ are denoted by $\mathrm{Re}(\mathbf{A})$ and $\mathrm{Im}(\mathbf{A})$. $A[k, \ell]$ denotes the entry of $\mathbf{A}$ in the $k^{\text{th}}$ row and the $\ell^{\text{th}}$ column. $a_{\ell}$ refers to the $\ell^{\text{th}}$ element of $\mathbf{a}$ and $\mathbf{a}_{\ell}$ refers to the $\ell^{\text{th}}$ column of $\mathbf{A}$. Similarly, $\mathbf{A}[:, k]$ refers to the $k^{\text{th}}$ column of $\mathbf{A}$. Unspecified norm equations are $\|\mathbf{a}\|_{2} = \mathbf{a}^{*}\mathbf{a}$ for vectors, and the Frobenius norm $\|\mathbf{A}\|_{F} = \sqrt{\mathrm{Tr}(\mathbf{A}\mathbf{A}^{*})}$ for matrices. We define $j = \sqrt{-1}$. $\lfloor a \rfloor$ is the nearest integer from truncating $a$ and $\lceil a \rceil$ is the nearest integer greater than or equal $a$.

## II. System model

We begin by defining the classical representation of a received sample set, comprised of multiple sinusoids, in noise. We then clarify how quantization and normalization are performed, based on common hardware considerations for gain control. Afterwards, we summarize the learning objects of each subtask and present two loss functions based on prior work and the role of our algorithm in processing signals.

An $m$ multi-sinusoidal signal is represented by vectors of amplitudes, phases, and frequencies $\mathbf{a}, \boldsymbol{\phi}, \mathbf{f} \in \mathbb{R}^M$. The sinusoidal components are summed and sampled at sampling intervals $T$ as

$$u[n] = \sum_{i=1}^{m} a_i \exp\left(j2\pi f_i nT + \phi_i\right). \tag{1}$$

Given an infinite number of samples, the exact signal components can be resolved from this model, so long as the sampling period is small enough such that the sampling theorem [32] is maintained. In realistic settings, the number of samples is finite and the desired signal is obstructed by noise $v[k]$, which is often modeled as additive, independent, and identically distributed (IID), Gaussian noise. For brevity of notation, we define the $m \times N$-dimensional matrix $\mathbf{G}$ such that $\mathbf{g}_i = \{2\pi f_i nT + \phi_i\}_{n=0}^{N-1}$. Now (1) can be represented as a finite set of $N$ real and imaginary components with complex-valued noise

$$\mathbf{u} = \sum_{i=1}^{m} a_i \left(\cos\left(\mathbf{g}_i\right) + j\sin\left(\mathbf{g}_i\right)\right) + \mathbf{v}. \tag{2}$$

This model aligns with classic [13]–[17] and recent [25] approaches. We now extend the system model to include quantization effects.

Prior work with quantization has generally considered the estimation of only a single sinusoid [28], [33], [34] or one-bit resolution [30], [31], [35], thus defining the quantization levels to maximize the dynamic range (and limit clipping) is straightforward. Modeling quantization effects for multiple bit quantizers with varying signals is not as simple, and is not done consistently in literature [6], [8], [28], [36]. Yet, defining the quantization levels in a realistic manner is crucial for the system model. As a result, we define the quantization model for unit-norm power, based on ideal traditional gain control hardware found in wireless systems. Specifically, the signal model is normalized and quantized according to a $b$ bit uniform quantizer $\mathbb{Q}_b$ as

$$\mathbf{z} = \mathbb{Q}_b\left(N \frac{\mathbf{u}}{\|\mathbf{u}\|^2}\right) \tag{3}$$

$$\mathbb{Q}_b(A + jB) = \mathbb{Q}_b(A) + j\mathbb{Q}_b(B) \tag{4}$$

$$\mathbb{Q}_b(A) = q_i \quad \text{if } A \in (q_{i-1}, q_i]. \tag{5}$$

The assumption of ideal gain control is a simplification, however, the nonlinear effects of a realistic gain controller would almost certainly be eclipsed by the subsequent quantization function. There are $2^b$ quantization levels $\mathbf{q} := \{q_i\}_{i=0}^{2^b}$ that can be output from $\mathbb{Q}_b$ covering $2^b + 1$ bins in the range of $[-1, 1]$. The output of the quantization function is assigned $q_i$ if the input is in the range $(q_{i-1}, q_i]$, and the first and last bins extend to $\pm\infty$. We are interested in low resolution settings, where $B \in \{1, 2, 3\}$. Higher resolutions are often not significantly different from 3 bit resolution with idealized gain control, as we assume in (3). We further restrict our system to the two edge cases, $B = \{1, 3\}$ bits to highlight the key differences between extremely low resolution (one-bit) and modestly low resolution (three-bit) sampling.

Complex number support is limited in many deep learning frameworks, so we vectorize the real and imaginary components to form the $\mathbb{R}^{N \times 2}$ received signal matrix as

$$\mathbf{X} = \left[\text{Re}(\mathbf{z}), \text{Im}(\mathbf{z})\right]^T. \tag{6}$$

It is noteworthy that an alternative approach could be to use a polar representation, or define complex layers that operate using complex-valued optimization [37], [38]. Such architectures are an interesting topic of future investigation.

Given the input to the system from (6), we mathematically summarize the goal of our neural network from a general learning perspective. First, we divide the problem into two tasks, detection and estimation, to be learned by two different neural network architectures. Detection is performed first. The goal is for our algorithm to learn the parameters $\mathbf{W}_1$ for a function $\beta_{\mathbf{W}_1}$ that relates a set of observations $\mathbf{X}$ to the true number of sinusoidal components $m$ in the original $\mathbf{u}$, based on a loss function $\text{L}_{\text{det}}$. The estimation network has parameters $\mathbf{W}_2$ and is trained to predict $\{\hat{\mathbf{a}}, \hat{\mathbf{f}}, \hat{\boldsymbol{\phi}}\}$, assuming perfect knowledge of $m$, with a loss function $\text{L}_{\text{est}}$. The training can be summarized as solving the problem

$$\mathbf{W}_1 = \underset{\mathbf{W}_i \in \mathbf{W}_{\text{det}}}{\arg\min} \mathbb{E}_{\mathbf{X}}[\text{L}_{\text{det}}(m, \beta_{\mathbf{W}_i}(\mathbf{X}))] \tag{7}$$

$$\mathbf{W}_2 = \underset{\mathbf{W}_i \in \mathbf{W}_{\text{est}}}{\arg\min} \mathbb{E}_{\mathbf{X}}[\text{L}_{\text{est}}(\{\mathbf{a}, \mathbf{f}, \boldsymbol{\phi}\}, \Theta_{m, \mathbf{W}_i}(\mathbf{X})) \,|\, m]. \tag{8}$$

The networks are combined and the overall estimate is produced according to

$$\hat{m} = \beta_{\mathbf{W}_1}(\mathbf{X}) \tag{9}$$

$$\{\hat{\mathbf{a}}, \hat{\mathbf{f}}, \hat{\boldsymbol{\phi}}\} = \Theta_{\hat{m}, \mathbf{W}_2}(\mathbf{X}). \tag{10}$$

All of the outputs in (10) are real vectors of size $\hat{m}$. Additionally, (10) and (8) imply that the estimator $\Theta_{\hat{m}}$ is specific to the value of $\hat{m}$ provided during detection. We make use of this in Section III, where internal reconstruction is used to estimate the $\hat{m}$ signal set.

We note that (7)-(10) can be applied to the traditional algorithms without quantization by reframing the training steps and meaning of $\mathbf{W}_1, \mathbf{W}_2$. By regarding the "learning" phase as a period of collecting sample statistics, (7) and (8) can be thought of as a tuning process for determining Bussgang gain [11] and algorithm parameters such as dither scaling [30]. The minimization is performed using known equations, such as the multi-level quantization presented in [8]. Then the functions $\beta, \theta$ can be regarded as the combination of the Bussgang Decomposition, a detection algorithm (e.g. AIC) and an estimation algorithm (non-parametric/parametric).

The goal of our network is to estimate the relevant sinusoidal components as an initial signal processing task, leaving possible filtering or decoding to subsequent algorithms. As a result, the detection algorithm should learn to overestimate the number of signals during uncertainty, rather than underestimate. This way higher level algorithms can filter out small or irrelevant components rather than miss signals altogether. To instill this knowledge, we define a heavy-sided loss function for the detection network as

$$L_{\det}(m, \hat{m}) = \begin{cases} e^{m-\hat{m}} - 1 & \text{if } m \geq \hat{m} \\ \frac{1}{2}(m - \hat{m})^2 & \text{otherwise.} \end{cases} \quad (11)$$

The specific internal functions are chosen to be differentiable, smooth, and ensuring that the loss has the relationship $L_{\det}(m, m+1) < L_{\det}(m, m-1) < L_{\det}(m, m+2)$, which we make use of in our analysis of the learning threshold. Note that differentiability and smoothness are determined by the functions, rather than the input space. We do not seek to fully define the bounds and assumptions implicit in this setting, but interested readers are encouraged to review statistical learning resources such as [26].

In the estimation module, we will train the network based on mean squared error (MSE), assuming the true number of signals m is known. Because the final model could potentially have different model order predictions, a similar loss function cannot be used for the overall SignalNet. Instead, when both modules are combined, we evaluate the network according to the Chamfer distance [39], which is used in [25] and defined as

$$L_{\text{chamfer}} = \sum_{f_i \in \mathbf{f}} \min_{\hat{f}_k \in \hat{\mathbf{f}}} |f_i - \hat{f}_k| + \sum_{\hat{f}_i \in \hat{\mathbf{f}}} \min_{f_k \in \mathbf{f}} |\hat{f}_i - f_k|. \quad (12)$$

The comparison can be applied to the frequency estimates as shown, or the other desired quantities. Effectively, this loss function penalizes for detecting the incorrect number of signals $\hat{m} \neq m$ as well as for incorrect estimates. Poor estimates are penalized twice, however, due to the two summations. An alternative loss function could be based on cosine similarity. We believe that the correct approach is to use the Chamfer distance between the vectors of different sizes. The reason is that the parameter values can not necessarily be chosen to indicate model order. For example, zeros in the amplitude vector indicate the absence of a sinusoid while while zeros in the phase vector imply the presence of a sinusoid with zero phase

## III. SignalNet

In this section, we define the SignalNet architecture, comprised of two sub-networks for detection and estimation. Readers interested in learning more about the basics of deep learning are encouraged to review [40], [41]. We focus heavily on the estimation side of the problem, because extraneous signals can be eliminated at later points, but poorly estimated signals are detrimental to application-specific information. To improve the learning of our estimation module, we include domain knowledge through reconstruction and cancellation, similar to successive interference cancellation [42], with similar algorithms also based on it such as [43]. We provide a description, figures, and a table summarizing the operation of SignalNet, as well as source code[1].

We design SignalNet, shown by the modules and overall architecture in Figures 1-3, according to the general structure in (6)-(10), where one module detects the number of sinusoids in the signal and another estimates the sinusoids parameter from an observed sequence $\mathbf{X}$. The detection module is used to estimate the number of sinusoids present in a signal, and is modeled using a three layer convolutional neural network with a softmax output. We use a softmax activation function on the output, which outputs a one-hot vector representing the most likely number of sinusoids detected. Although a real-valued system could be used with rounding, the domain of the output is significantly larger (i.e. the outputs are no longer a discrete set of $M$ values), causing the performance to be negatively impacted.

The sinusoid estimation module, however, requires more direct knowledge of the relationship between the output parameters and the input $\mathbf{X}$ to effectively capture the information. To do this, we design the estimator around the idea of successive-estimation and cancellation, similar to interference cancellation methods in non-orthogonal communications [42]. The network architecture first estimates the parameters for a single sinusoid, reconstructs the time-domain signal, and compares it with the original input. Then, it estimates the parameters for the next sinusoid from the difference. This knowledge helps our neural architecture better handle multiple frequencies and is intuitive. A block diagram of the sinusoid estimator is shown in Figure 2. Internally, each sinusoid estimator has two branches of two convolutional layers where one branch has batch normalization for estimating frequency and phase components and the other does not to retain amplitude information. Hyperparameters are tuned using iterative grid search for both the sinusoid estimator and detection module, and layer specifics can be found in Table I.

After completing the model tuning, we test against two traditional architectures, a multilayer perceptron (MLP) and a convolutional neural network (Conv). Each of the two basic networks are constructed with a similar number of parameters and trained in the same way as our network. The comparison networks are comprised of two hidden layers, and end with the same number of outputs corresponding to the amplitude, frequency, and phase as our sinusoidal estimator. We use this to show that our novel architecture is able to learn more effectively than a simple network of similar size. The results of this comparison are shown in Figure 8 in Section VIII.

---

[1]https://github.com/Ryandry1st/Carrier-Frequency-Offset

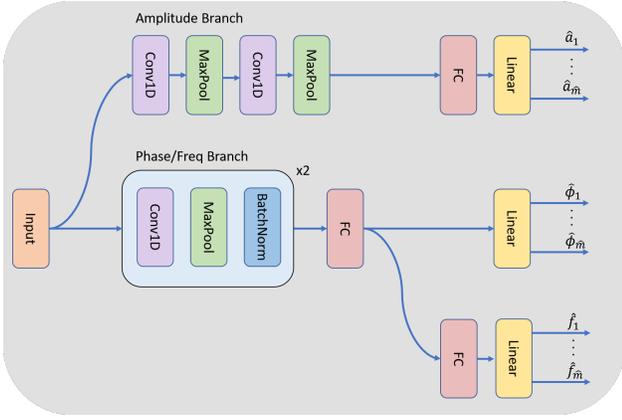

Fig. 1. A diagram of a block estimator used to determine the parameters for a specific number of sinusoids. The two path network is made up of an unnormalized branch for amplitude estimation and a normalized branch for frequency and phase estimation. Frequency and phase estimation are linked because of the natural linear relationship between frequency and change in phase over a set of $N$ samples. This is equivalent to determining the frequency as the average phase change over the $N$ samples.

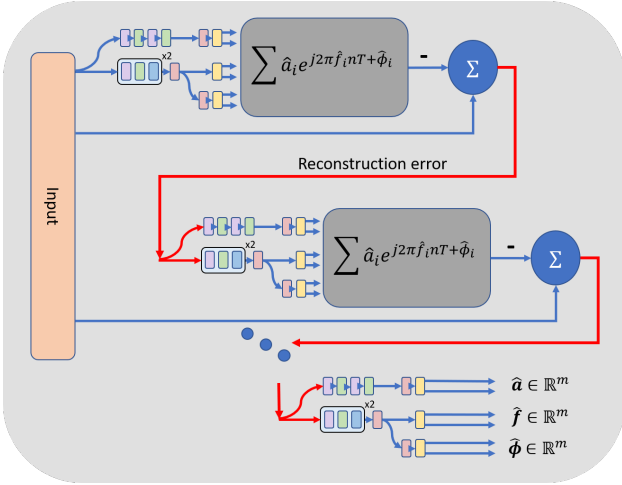

Fig. 2. The sinusoid estimator is comprised of multiple block estimators. These networks successively estimate $\{1, \dots m\}$ sinusoids by estimating and reconstructing each number of sinusoids and recomputing on the error. Internal reconstruction is used to provide clear relationships between the estimated outputs $[\mathbf{a}, \mathbf{f}, \boldsymbol{\phi}]$ and the input $\mathbf{X}$. This formulation results in each $m \in 1, \dots M$ estimator producing different outputs and learning different features.

Recalling (10), we train all of the networks independently, where each $m \in \{1, \dots M\}$ possible number of outputs and each quantization resolution $b \in \{1, \dots B\}$ result in a different network, totaling $M \times B$ networks in our investigation. In this setup, we build the a $b$-bit SignalNet architecture from $M$ different sinusoid estimator networks. The need for having separate estimators for each number of sinusoids is based on two considerations: real-time feasibility and our choice of internal reconstruction. The real-time feasibility is a result of graph-based optimiza-tion, which does not allow for dynamically-sized outputs without sacrificing inference speed. Offline training is not considered in this paper, as our algorithm only needs to be trained and then deployed. Along with computational performance, our use of internal reconstruction requires separate estimators for each number of sinusoids. When estimating and reconstructing an unknown number of sinusoids, the first sinusoid estimated will be different depending on the total number of sinusoids because of the goal of reconstructing a similar set of samples as the network's input. In other words, our network does not attempt to find the peaks of DFT bins. Instead, it finds the parameters that generate a signal most closely approximating the input, while gradients are only updated from the parameter error. This is best understood from the visualization in Fig. 4, where we show a two-sinusoid signal with the best reconstruction using $\{1, 2, 3\}$ sinusoids. The output of the single sinusoid estimator will not be either of the two underlying sinusoids, but some sinusoid between the two. After training, the corresponding detection and estimation modules are joined according to Figure 3.

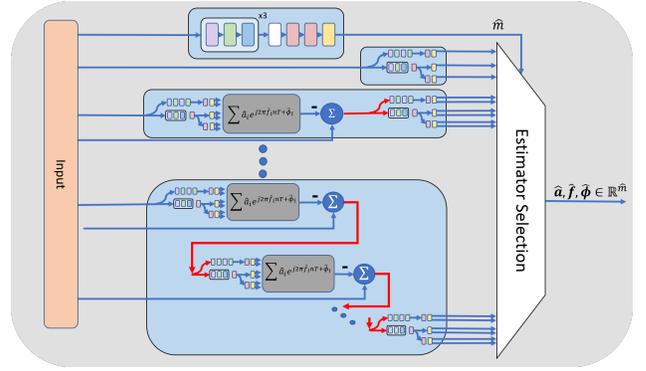

Fig. 3. A block diagram of SignalNet. The architecture includes $M$ sinusoid estimator modules and one detection module, all trained for a specific quantization resolution $b$. Each subnetwork is highlighted, with the top network being the detection module, and the following networks are the $\{1, 2, \text{and} M\}$ sinusoid estimators. In our investigation, we develop a SignalNet variant for one-bit and three-bit resolutions, resulting in $(M + 1) \times 2$ total networks. Distillation [44] could also be considered to overcome the need for specific variants, but we leave this for future work.

## IV. Simulation setup

We consider a simulation setup similar to [25], with the notable exception of considering a reduced range of sinusoids $M = 5$. The most important aspect is the non-uniform frequency distribution of our data generation, as outlined in Algorithm 1. The frequency generation is unusual by necessity due to limitations on the frequency spacing and to remove unintended bias from the network performance. One might consider using frequency components drawn from an independently sampled uniform distribution. Unfortunately, it has been shown that resolving frequency content with spacing equal or less than $1/N$ from $N$ discrete, noisy samples is non-convergent and may result in excessive errors [45], [46]. Although off-grid techniques can be used to resolve sparse signals (for some appropriate basis), resolving two signals with less separation and in the presence of noise is provably inconsistent [46]. In contrast, we could use uniformly spaced frequency content, however, this would introduce significantly more structure to the data that would unintentionally bias the neural network. For example, rather than estimate each sinusoid, the network would simply need to learn to estimate the spacing and a single frequency with high accuracy, thereby changing the structure and goal of the system. A neural network trained in such a way would be unlikely to generalize well to less structured settings.

TABLE I
Network parameters for one detection module and sinusoid estimator.

| Network | Layer | Primary Parameter | Activation Function | Output Dimension |
|---|---|---|---|---|
| Detection Module | Conv + Pooling + BN | 32 Filters | ReLU | (32, 32, 32) |
| Detection Module | Conv + Pooling + BN | 64 Filters | ReLU | (32, 16, 64) |
| Detection Module | Conv + Pooling + BN | 128 Filters | ReLU | (32, 4, 128) |
| Detection Module | Dropout | 0.7 Rate | | (32, 512) |
| Detection Module | Fully Connected | 128 Neurons | ReLU | (32, 128) |
| Detection Module | Fully Connected | 64 Neurons | ReLU | (32, 64) |
| Detection Module | Fully Connected | 5 Neurons | Softmax | (32, 5) |
| Sinusoid Estimator | Conv + Pooling + BN | 8 Filters | ReLU | (32, 32, 8) |
| Sinusoid Estimator | Conv + Pooling + BN | 16 Filters | ReLU | (32, 16, 16) |
| Sinusoid Estimator | Fully Connected | 16 Neurons | SeLU | (32, 16) |
| Sinusoid Estimator | Fully Connected | $\hat{m}$ Neurons | | (32, $\hat{m}$) |
| Sinusoid Estimator | Conv + Pooling | 8 Filters | ReLU | (32, 32, 8) |
| Sinusoid Estimator | Conv + Pooling | 16 Filters | ReLU | (32, 16, 16) |
| Sinusoid Estimator | Fully Connected | 16 Neurons | SeLU | (32, 16) |
| Sinusoid Estimator | Fully Connected | $\hat{m}$ Neurons | | (32, $\hat{m}$) |
| Sinusoid Estimator | Conv + Pooling | 16 Filters | ReLU | (32, 16, 16) |
| Sinusoid Estimator | Fully Connected | 16 Neurons | SeLU | (32, 16) |
| Sinusoid Estimator | Fully Connected | $\hat{m}$ Neurons | | (32, $\hat{m}$) |

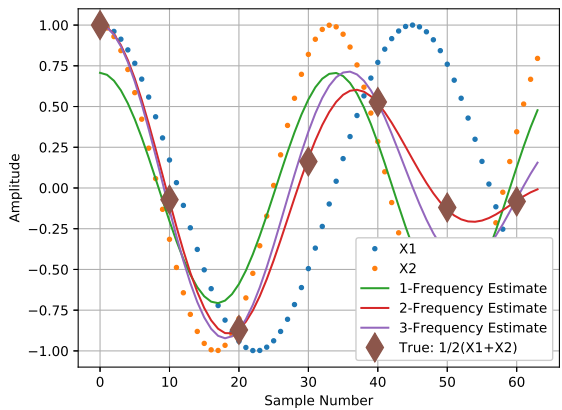

Fig. 4. An illustration of the internal reconstruction task using values from the estimates of SignalNet. Although the network is trained based on the mean squared estimation error, the internal layers seek to most closely represent the signal for a given $\hat{m}$. In the figure, neither X1 nor X2 are output when using only one sinusoid, shown in green. Instead, the best estimate is that which reconstructs the sinusoid created by the brown diamonds.

Instead, we use a similar setting to [25], where we start by selecting a random number of sinusoids to be present. Then the uncertainty in the spacing is selected by a folded normal distribution along with the initial frequency offset from a uniform distribution. Finally, the maximum frequency is compared with 0.5, to ensure that no frequency content is undersampled according to the Nyquist criterion [32]. The frequency distribution for each m is shown in Fig. 5. Then amplitudes and phases are drawn from uniform distributions and the signal is constructed by summing over all of the sinusoids and adding Gaussian noise to obtain $\{y\}_{n=0}^{N-1}$ for a desired SNR. In the final steps, the signal is normalized to unity power norm, representative of an ideal automatic gain controller or similar control systems at the input to the data converter. Finally, quantization $\mathbb{Q}_b$ is applied according to the number of bits $b$.

We select the simulation setup based on two considerations for the difficulty (i.e. challenge in achieving accurate estimates) of the problem: the length of the data and the separation of the frequencies. For length $N$ discrete observations, resolving unquantized data with separation of $1/N$ is nearly impossible in the presence of noise [45], [46], and in particular when the SQNR is below 20 [47] as we have here. We do not strictly enforce this separation, but use the folded normal distribution to discourage it, resulting in some situations that are nearly unresolvable. This leads to estimators that may have overlapping frequency content that could represent interference. Additionally, the amplitude of the signal directly contributes to the estimation error of the problem, so we constrain the amplitudes to the range of [0.1, 1.0]. These choices create a challenging simulation setup, as seen by the simulation results in Section VIII, especially the amplitude estimation results in Figure 10.

When testing the results, the data is drawn again from the same distributions according to Algorithm 1. This is not the same as if the training data is used for testing, only that the distributions the values are drawn from are

## Algorithm 1 Simulation Data Generation

1: Draw a random integer, $m$ from $[0, \ M-1]$
2: **do**
3:     $w_0 \sim \mathcal{U}(0, 0.25)$
4:     $f_1 \leftarrow w_0$
5:     **for** $i \in [1, ...m-1]$ **do**
6:         $w_i \sim |\mathcal{N}(0, 2.5/N)|$
7:         $f_{i+1} \leftarrow w_0 + i/N + w_i$
8:     **end for**
9: **while** max $f_i > 0.5$
10: $a_i \sim \mathcal{U}(0.1, 1.0) \quad i \in [1, ...m]$
11: $\phi_i \sim \mathcal{U}(0, 2\pi) \quad i \in [1, ...m]$
12: $\mathbf{y} \leftarrow \sum_{i=0}^{m} a_i \exp(2\pi j f_i n + \phi_i) + \mathbf{v}$
13: $y_i \leftarrow \frac{y_i}{\|\mathbf{y}\|} \quad i \in \{0, 1, ..N-1\}$
14: $\mathbf{x}_i \leftarrow \mathbb{Q}_b(y_i)$

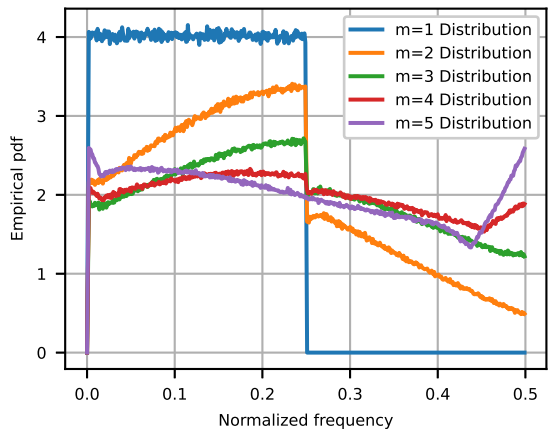

Fig. 5. Distribution of frequency content **f** over 10 million trials. The density functions show how the changing value of m strongly affects the frequency distribution, causing significantly more challenging results as m increases. It is clear this distribution, with knowledge of m, contains information that can be leveraged by our neural network.

the same, which is not a particularly strict assumption considering train-test splits are often done so that the distributions remain the same. That being said, we also evaluate the case where the frequencies are drawn from a different distribution in Section VIII-G.

## V. Benchmarks

Because there are, to the best of our knowledge, no other algorithms designed for sinusoidal decomposition and estimation from quantized data, we limit the benchmarks to well-established estimators, in combination with Bussgang Decomposition [11]. We first provide a brief recount of Bussgang Decomposition and how it applies to our system model. Afterwards, we highlight the two common detection methods, AIC and MDL and the limitations that minimizing model order has on the detection loss function, (11). An alternative method, BIC [48] also has potential, though it is known to penalize over-estimation even more than AIC, which will inherently perform worse for our loss function, and can be shown to have equal performance to

MDL in our scenario [26, pg. 236]. We also investigate EM-VAMP, a message passing algorithm, as a method for determining the model order in a compressed sensing form. EM-VAMP is an excellent bridge between traditional methods like AIC, and data driven methods like deep learning. In fact, AMP-based methods have been shown to outperform neural networks in similar scenarios such as low-resolution channel estimation [49]. Finally, we suggest the non-parametric methods we compare against based on the Fast Fourier Transform (FFT) and Periodogram.

The Bussgang Decomposition is a method for linearizing a function by computing the linear minimum mean squared error estimator. In the case of quantization, it has been shown to be a useful tool to separate the signal and quantization noise into two uncorrelated terms [8], [11], [50] assuming a gaussian signal. The essence of the decomposition is to calculate a gain factor $G$ such that, for a nonlinear function $y = U(x)$, the result is

$$y = Gx + \eta$$
$$\mathbb{E}[y\eta] = 0$$
$$\mathbb{E}[x\eta] = 0$$

which is the linear minimum mean squared error estimator (LMMSE) of $x$ from $y$. This formulation extends to vector signals, causing $G$ to become $\mathbf{G}$ which is a diagonal matrix. The exact formulation for multi-bit quantization, $U = \mathbb{Q}_b$, is provided in [8]. One of the key components in the formulation is knowledge of the first and second moments of $\mathbf{x}$. Because of our power normalization and gaussian noise, we observe that the complex observation vector $\mathbf{X}[:, 0] + j\mathbf{X}[:, 1] \sim \mathbb{CN}(\mathbf{0}, \mathbf{1})$. Then, by using the inverse of $\mathbf{G}$ and neglecting the distortion, we can obtain the linearized recovery of $\mathbf{x}$

$$\tilde{\mathbf{x}} = \mathbf{G}^{-1}(\mathbf{X}[:, 0] + j\mathbf{X}[:, 1]). \tag{13}$$

We will apply this decomposition to the following benchmark detection and estimation algorithms in our simulation results.

We evaluate our detection module against AIC and MDL, although both of these methods are suboptimal due to misaligned goals. Fundamentally, these methods attempt to minimize the complexity while capturing the number of spectral components. In contrast, the loss function (11), penalizes under-estimates more than over-estimates. This, in part, leads to significantly under-performing detection results for AIC and MDL. Additionally, AIC and MDL are not well-suited for low resolution sampling, so in the case of 1-bit data, the results are non-convergent and have loss values far greater than other algorithms. To keep the scaling of the graphs meaningful, we instead show the 2- and 3- bit results for these two algorithms. We also show the EM-VAMP method, using a learned threshold for determining the model order from the recovered vector support. While not following the exact same structure as we outlined in Section II, this benchmark is similar to data driven algorithms, but is more structured than neural networks. The EM-VAMP method recovers the frequency-grid-aligned sparse signal **x** (the sinusoidal components in the frequency domain) from a 2x oversampled DFT observation matrix **A**. Mathematically, the EM-VAMP method works to minimize the error from observations **y** for the model

$$\mathbf{y} = \text{sign}(\mathbf{A}\,\mathbf{x} + w) \tag{14}$$

by recovering the sinusoidal amplitudes and phases in the sparse coefficients of **x**. This framework is limited due to its on-grid nature, however, so we only consider it for detection. For more information on EM-VAMP and GAMP methods, see [23], [24].

To benchmark our multi-sinusoid estimator, we employ the Periodogram, because it has been shown to produce better estimates for low resolution sinusoidal data than eigendecomposition and dithering methods [12], [30]. The Periodogram is calculated from the scaled-and-squared, zero-padded Fast Fourier Transform (FFT) with $N_0 = 2^{16}$ to ensure that grid resolution is not a limiting factor. Although amplitude and frequency information can be obtained from the Periodogram, the phase information is recovered from the inverse tangent of the ratio between the imaginary and real components of the FFT. Precisely, the phase estimate is calculated first by the FFT of the samples, then selecting the $m$-largest magnitude of the FFT, and finally applying the inverse tangent (this would be atan2 in many programming languages). The steps can be mathematically expressed as

$$\mathbf{r} = \text{FFT}([\mathbf{x}, \mathbf{0}_{N_{\text{FFT}}-N}]) \in \mathbb{C}^{N_{\text{FFT}}} \tag{15}$$

$$\mathbf{p} = m - \text{argmax}(\mathbf{r}) \tag{16}$$

$$\equiv \{i \text{ for } p_i \in \text{abs}(\mathbf{r}) : |\text{abs}(\mathbf{r}) \cap (-\infty, p_i]| \leq m\} \tag{17}$$

$$\hat{\boldsymbol{\phi}}_{\text{FFT}} = \arctan\Big(\text{Im}(\mathbf{r}[\mathbf{p}]), \text{Re}(\mathbf{r}[\mathbf{P}])\Big). \tag{18}$$

We note two important clarifications in (15)-(18). First, the peak finding algorithm in step 2 simply finds the $m$ largest maxima of the oversampled $N_{\text{FFT}}$ FFT. Second, we make use of both the absolute value and the size of a vector in the definitions, so we use $abs()$ to refer to the magnitude of the complex values and $|\ldots|$ to refer to the cardinality of the set.

## VI. Learning Threshold

Before looking at the simulation results, we first introduce a baseline for algorithmic comparison, particularly for neural network evaluation. Because machine learning algorithms learn from data, there are inherent constraints and underlying distributions that algorithms can optimize for, creating unfair bias when compared to traditional algorithms. For this reason, we also define a learning threshold based on statistical estimation theory and Empirical Risk Minimization [26]. First, let $\mathbf{a} \in \mathbb{R}^p$ be the input and $b \in \mathbb{R}$ be the true output, with joint distribution $P(\mathbf{a}, b)$. The goal is for an estimator, $g_\theta(\mathbf{a})$ that is defined by parameters $\theta$, to predict $b$. We wish to minimize a loss function, and for simplicity we start with mean squared error (MSE), for $K$ predictions. Note, sample-wise MSE is defined as $\hat{L}_{\text{MSE}}(b, g_\theta(\mathbf{a})) = 1/K \sum_{i=1}^{K}(b_i - g_\theta(\mathbf{a}_i))^2$, where the parameters $\theta$ are updated based on training data. If the dataset is representative of the distribution, which occurs for sufficiently large numbers of samples in expectation, the empirical formulation matches the theoretical formulation as

$$\min_\theta L_{\text{MSE}}(b, g_\theta(\mathbf{a})) = \min_\theta \mathbb{E}_{\mathbf{a}} \, \mathbb{E}_{b|\mathbf{a}}[(y - g_\theta(\mathbf{a}))^2 \,|\, \mathbf{a}] \tag{19}$$

$$g_\theta(\mathbf{a}) = \mathbb{E}[b \,|\, \mathbf{a}]. \tag{20}$$

Now, we define the learning threshold as the worst-case training error, in the non-adversarial case, that occurs when **a** is independent from $b$. That is, when **a** is both independent from $b$ and not chosen by an adversary to intentionally mislead an algorithm to the wrong outcome. This simplifies (19)-(20) to

$$L_{\text{threshold}}(b, g_\theta(\mathbf{a})) = \min_\theta \mathbb{E}_b[(b - g_\theta(\mathbf{a}))^2] \tag{21}$$

$$g_\theta(\mathbf{a}) = \mathbb{E}[b] \quad \forall \, \mathbf{a}. \tag{22}$$

The result, for this trivial example, is that the estimator should simply estimate the mean of the output variable, and the resulting error will be the variance of $b$. This well-known relationship is a foundational concept of our learning threshold, and is directly used for our algorithms trained with MSE loss. It is similar to other established estimation techniques such as the minimum variance unbiased estimator or the minimum mean squared error estimator when the loss function is the mean squared error as we have shown. The formulation extends beyond mean squared error though, and can be applied to more interesting cases, in particular the detection module with loss function defined by (11). This distribution is heavy-sided and the output variable can only take integer values, leading to more extensive analysis.

The threshold analysis is not just an interesting derivation. We show in Section VIII that our algorithms for both detection and amplitude estimation converge to this threshold for one-bit data. This implies that the learning done by the neural networks is limited to distributional learning, which is simply learning to minimize the loss for an output variable's distribution and loss function. The effective result is equivalent in performance to a constant output function that simply estimates the loss-aware average. Therefore, we can also see the inherent limitations of the threshold estimator: the output distribution and loss function must be known, and the constant estimator is only reasonable in expectation, with potentially poor results. It is only truly useful to judge whether a basic level of learning has occurred, rather than as a true estimator.

Furthermore, the learning threshold can be extended to vector valued outputs by calculating the expected loss incurred as a result of applying the threshold estimator to each scalar output. This is equivalent to determining the constant estimator for each output and then evaluating the resulting expected loss, exactly as if the loss function were applied to any general estimator. It is important to note that knowledge of different sinusoidal estimates does not improve the learning threshold because each of the sinusoidal parameters come from independent realizations and there is no observation that could relate one set of sinusoidal parameters to another.

The learning threshold will show that there is an inherent bias (in a learning theory sense) that is observable just from the distributions and loss function that can often surpass traditional methods. The threshold is relevant for two reasons: 1) it provides a scalar metric for the uncertainty in the output variable, with respect to the loss function, and 2) comparing the neural network's loss with this threshold provides clarity about whether a neural network is learning features or blindly optimizing for distributions. In this case, blind optimization refers to equal performance to the learning threshold, which gives the same constant outputs regardless of the inputs. In general, we expect our neural network estimators to outperform the learning threshold, suggesting some relationship is learned between the input and the output. Because the quantization effects are not independent from the noise however, low SNR data can be misleading or adversarial. In contrast, the threshold does not change with the input SNR, so we would expect to see results that are similar to or worse than the learning threshold for sufficiently low signal to noise ratios and quantized signals. In the subsequent sections we will derive learning thresholds for the detection and estimation problems based on our simulation setting to lend some intuition to why neural networks outperform traditional methods. In the final section, we will evaluate our network using data from a different distribution, showing the generalizability of our model and the relationship to the learning threshold.

## VII. Model training

In our setup, we train the two modules, the detection module and the sinusoid estimation module, separately. We employ the loss function in (11) to train the detection module for 20 epochs with $50,000$ realizations of one-hot encoded signal counts. We evaluate the estimation module using the MSE in the training data, assuming that the true number of signals is known. This way each amplitude-frequency-phase estimate corresponds to a true set of parameters. Note that the mean-squared-error is defined for multi-output systems as the mean over the stacked vector of outputs, which is

$$L_{\text{MSE}}(\mathbf{c}, \hat{\mathbf{c}}) = \frac{1}{p} \| \mathbf{c} - \hat{\mathbf{c}} \|^2 \quad \mathbf{c}, \hat{\mathbf{c}} \in \mathbb{R}^p. \tag{23}$$

Unfortunately, not all parameters have the same scale or degree of randomness, relative to the loss function. Classically, this is mitigated by scaling all of the sample outputs to roughly mimic a standard normal distribution, or to be within the same range i.e. $[0, 1]$ [51]. In our case, because the output variables are not drawn from the same distribution, but the distribution is known, this technique does not appropriately scale the outputs. The learning threshold, as we defined in Section VI, is used to solve this problem by measuring the loss function with respect to the distribution of the output variables. We define the overall normalization used to achieve a unified scalar loss function as

$$\boldsymbol{\ell} = \left[ L(\mathbf{a}, \hat{\mathbf{a}}), L(\mathbf{f}, \hat{\mathbf{f}}), L(\boldsymbol{\phi}, \hat{\boldsymbol{\phi}}) \right]^T \tag{24}$$

$$\boldsymbol{\ell}_{\text{threshold}} = \left[ \frac{1}{L_{\text{threshold},\mathbf{a}}}, \frac{1}{L_{\text{threshold},\mathbf{f}}}, \frac{1}{L_{\text{threshold},\boldsymbol{\phi}}} \right]^T \tag{25}$$

$$\boldsymbol{\ell}_{\text{eff}} = \frac{1}{m} \boldsymbol{\ell}^T \boldsymbol{\ell}_{\text{threshold}}. \tag{26}$$

In this case, each of the respective estimated parameters is stacked into a vector, and compared to the true parameters in mean squared error and represented by $L(\mathbf{x}, \hat{\mathbf{x}})$. Then the scaling (26) is applied to prevent the gradients from being dominated by parameters that have larger variance i.e. the true phase values are drawn from a larger range than the amplitude or frequencies. We then train on $100,000$ data samples, using gradient updates from a scaled sum of loss functions based on the learning thresholds. We also use (26) again for computing the overall chamfer loss as a unified metric for SignalNet. The specific learning thresholds are derived in Section VIII. While training, we use learning rate reduction and early stopping based on a separate validation set to optimize the gradient learning. Batch sizes are set to 32, and networks are all trained using an Adam optimizer [51], with an initial learning rate of $0.001$. When evaluating our algorithms, we generate $8,000$ samples of new data following the same algorithm for our test data.

Before evaluating our algorithm, we briefly remark on data size and memorization. The size of the training dataset, while initially appearing large, is extremely small versus the input data space. For example, in the smallest case, $b = 1$, the size of the input data space is $2^{2N}$, where $N = 64$ and there is a real and imaginary component, each taking a value of $\pm 1$. Similarly, the size of our neural networks is much smaller than the data dimension as well, with the total number of parameters for SignalNet being $300,000$ parameters, but only at most $120,000$ parameters in any individual network. As a result, our algorithm is not capable of memorizing the data, and could potentially benefit from increasing the dataset size by orders of magnitude. Our results in Section VIII show the efficacy of our algorithm to learn under this strict data constraint. Furthermore, near real-time inference is achievable with modern processing hardware. With an Nvidia 3090 GPU and limited optimization, we were able to achieve inferences in under 1 ms.

## VIII. Simulation results

In this section, we evaluate each SignalNet component, first individually, then as a whole. We derive the learning threshold for each subtask, shown by dashed lines in each plot, and compare the simulation results of our algorithm along with the benchmark traditional algorithm. In the final setting, we combine the two modules and compare it with different combinations of our modules and the traditional methods.

### A. Detection results

First, we evaluate the detection module, which is compared with EM-VAMP, AIC and MDL algorithms in Fig. 6. In this situation, the loss function is the detection loss defined in (11), and the possible outputs are $\hat{m} \in \{1, 2, ..5\}$. Then, the learning threshold is defined as

$$L_{\text{threshold,m}}(m, g_\theta(\mathbf{X})) = \min_{\hat{m}} \mathbb{E}_m[L_{\text{det}}(m, \hat{m})] \tag{27}$$

where $\hat{m} = g_\theta(\mathbf{X}) \ \forall \ \mathbf{X}$ is a constant estimator. Because the function is smooth, differentiable, and convex, it is a simple step to go from (27) to solving for $\hat{m}$ from

$$0 = \frac{\partial}{\partial m} \mathbb{E}_m[L_{\text{det}}(m, \hat{m})]. \tag{28}$$

Then, because $L_{\text{det}}$ is bounded over the inputs, the derivative and expectation can be interchanged by the bounded convergence theorem. Following from (28),

$$0 = \mathbb{E}_m \left[ \frac{\partial}{\partial m} L_{\text{det}}(m, \hat{m}) \right] \tag{29}$$

$$= \frac{1}{M} \sum_{m=1}^{M} \frac{\partial}{\partial m} L_{\text{det}}(m, \hat{m}). \tag{30}$$

The step from (29) to (30) is because m is uniformly distributed and discrete, so the expectation is the arithmetic mean. In our definition of AIC, we made the important restriction that $L_{\text{det}}(m, m+1) < L_{\text{det}}(m, m-1) < L_{\text{det}}(m, m+2)$, so the resulting threshold estimator will be $\hat{m} \in (\lceil M/2 \rceil, \lceil M/2 + 1 \rceil)$ for our simulation setting. We now ignore the constant $1/M$ factor, replace $\frac{\partial}{\partial m} L_{\text{det}}$ with its derivative components, and solve for the threshold estimator

$$0 = \sum_{m=1}^{\lfloor \hat{m} \rfloor} (m - \hat{m}) + \sum_{m=\lceil \hat{m} \rceil}^{M} e^{m - \hat{m}} \tag{31}$$

$$= \frac{1}{\lfloor \hat{m} \rfloor} \sum_{m=1}^{\lfloor \hat{m} \rfloor} (m) - \hat{m} + \frac{1}{\lfloor \hat{m} \rfloor} \sum_{m=\lceil \hat{m} \rceil}^{M} e^{m - \hat{m}} \tag{32}$$

$$= \frac{\lfloor \hat{m} \rfloor + 1}{2} - \hat{m} + \frac{1}{\lfloor \hat{m} \rfloor} \sum_{m=\lceil \hat{m} \rceil}^{M} e^{m - \hat{m}}. \tag{33}$$

Defining $\alpha = \frac{\lfloor \hat{m} \rfloor + 1}{2}$ simplifies the results to achieve a expression for the constant estimator:

$$0 = \alpha - \hat{m} + \frac{e^{-\hat{m}}}{\lfloor \hat{m} \rfloor} \sum_{m=\lceil \hat{m} \rceil}^{M} e^m \tag{34}$$

$$= -e^{\hat{m}}(\hat{m} - \alpha) + \frac{1}{\lfloor \hat{m} \rfloor} \sum_{m=\lceil \hat{m} \rceil}^{M} e^m \tag{35}$$

$$= -e^{\hat{m} - \alpha}(\hat{m} - \alpha) + \frac{1}{\lfloor \hat{m} \rfloor} \sum_{m=\lceil \hat{m} \rceil}^{M} e^{m - \alpha} \tag{36}$$

$$\hat{m} = W \left( \frac{1}{\lfloor \hat{m} \rfloor} \sum_{m=\lceil \hat{m} \rceil}^{M} e^{m - \alpha} \right) + \alpha \tag{37}$$

$$= W \left( \frac{1}{\lceil M/2 \rceil} \sum_{m=\lceil M/2 \rceil + 1}^{M} e^{m - \alpha} \right) + \alpha. \tag{38}$$

Here, $W()$ is the Lambert W function, and the final expression comes from the choice of $L_{\text{det}}$ and $\hat{m} \in (\lceil M/2 \rceil, \lceil M/2 + 1 \rceil)$. Evaluating for $M = 5$ leads to $\hat{m} \approx 3.69$, $L_{\text{threshold,m}} \approx 1.67$. We note that although the definition of m and $\hat{m}$ must be integer values, fractional amounts can be effectively obtained by choosing between the floor and ceiling values with the appropriate regularity. In this setting, the threshold estimator is not truly constant, but the estimator does not depend on the input. The threshold is shown in Figure 6 by the dashed line and provides a metric for the maximum expected error if the noise and quantization effects are negligible.

Evaluating the estimators, we first start with the simplest consideration: can the estimators perform better than a blind estimator? Most importantly, all of the estimators fail to reach the threshold for SNR $< -5$dB, showing that the noise and quantization effects cause the algorithms to perform worse than if the data $\mathbf{X}$ was ignored. Additionally, the one-bit detection module is never able to surpass the threshold, suggesting that the multiple sinusoid detection algorithm is not able to learn meaningful results from one-bit data. The three bit results are able to improve upon the threshold for SNR $> -4$dB, and perform better than higher resolution versions of AIC and MDL across the entire SNR range. Further, it can be observed that extremely low SNR data with greater variation (higher bit resolution) results in worse estimates. This validates the intuition that low SNR, coarsely quantized data is effectively adversarial compared to higher SNR data.

We can see that the neural network algorithms show a significant advantage over the traditional algorithms. It might be assumed this is due to the choice of loss function, which is different than the benchmark algorithms are designed for. This is not the case, however, as we show the results using simple mean-squared-error in Figure 7. While our algorithm's performance is not entirely due to the choice of loss function, we have already seen that the learning threshold is quite low, so we know that a simple estimator that estimates $m = 3, 4$ with the correct regularity will produce quite good results. That is the concept of distributional learning: where a sufficient amount of data with labels and corresponding loss values can lead to good neural network performance, so long as the data distribution is constant, and yet very little information from the inputs is used. In contrast, learning that is generalizable, i.e. learning features from the inputs, would apply even for other distributions.

### B. Architecture Comparison

We briefly demonstrate the effectiveness of our model architecture compared to traditional multi-layer neural networks. In this setup, we train and compare the proposed 3 bit sinusoidal estimator with m = 5 sinusoids against an equally sized multi-layer perceptron (MLP) and convolutional network (Conv). Training is performed in the same way, with the same datasets, and tested with 10,000 samples of data drawn from the same distribution. We show the comparison across all three estimation parameters in Fig. 8. We can see that our architecture is able to learn meaningful features and achieve the best results universally for SNR > −4 dB. The most obvious advantage our model has is in amplitude estimation, which achieves nearly 10 dB improvement and successfully improves with higher SNR. From these results we can see that our algorithm achieves more effective learning, with an equal number of parameters and data, compared to more straightforward approaches.

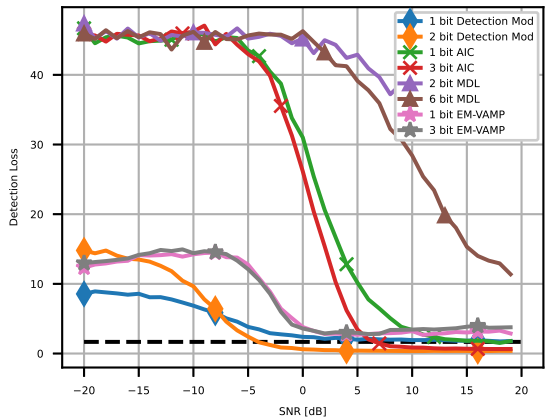

Fig. 6. Our detection module compared to EM-VAMP, MDL and AIC with the proposed detection loss metric. We do not show the results of 1-bit data resolution for MDL and AIC, due to a lack of convergence, which cause the detection loss to be too high and makes it harder to see the differences between the other performance results. Instead, we show the 2 and 3 bit results for comparison. It can be seen that the one-bit detection module is unable to surpass the learning threshold. The threshold is also far below either MDL or AIC until 7dB, showing the inherent advantage data driven techniques have, even for relatively uninformative distributions like a discrete uniform.

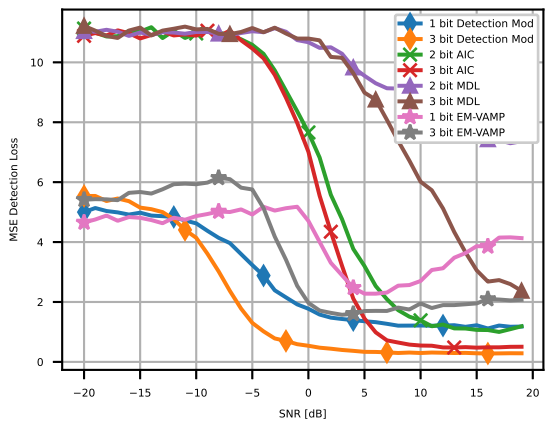

Fig. 7. A comparison of detection algorithms using traditional mean squared error (MSE). In many cases the results look similar to Figure 6, but we see that EM-VAMP is less competitive, compared to other algorithms, while MDL has become more competitive. Overall, our network performance for the detection problem remains similar, even with more general loss functions such as MSE.

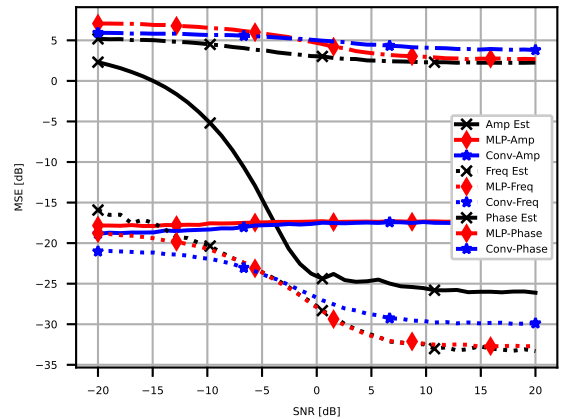

Fig. 8. A Comparison between our estimator architecture ({} Est), a mulitlayer perceptron (MLP-{}), and a convolutional network (Conv-{}) with three-bit data and m = 5. It can be seen that our architecture is the only one that successfully learns meaningful features for the amplitude estimation, while the other two models are unable to achieve meaningful improvements with increasing SNR.

### C. Frequency estimation

Next, we evaluate the frequency estimation performance of the estimation module. The distribution of the output vector $\mathbf{f}$ comes from both the initial $w_0$ and the offsets, $w_i$. While the compared algorithms must search for each point with no prior, the actual frequency distributions have structure that can be exploited for better estimation at low SNR. This is one reason why the Periodogram estimator performs significantly worse than the neural architecture in Figure 9, and does not reach the threshold for SNR $< -5$dB. Following Algorithm 1, the learning threshold is calculated for each number of sinusoidal components shown by assuming $m$ is known and computing the expected mean squared error. The estimator, $g_\theta(\mathbf{X})$ is simply the mean estimator because the loss function is the mean squared error, exactly as shown in Section VI. Note that the calculations are carried out in linear scale, although the plotting is done in dB scale.

First, we note the first and second moments of the folded normal distribution, used for the frequency components beyond the initial one, and approximated for our simulation settings

$$\mathbb{E}[w_i] = \sqrt{\frac{5}{N\pi}} \approx 0.1577 \quad i \in \{1, \ldots \text{ m}\} \tag{39}$$

$$\sigma_{w_i}^2 = \frac{2.5}{N} - E[w_i]^2 = \frac{5}{2N}(1 - \frac{2}{\pi}) \approx 0.0142. \tag{40}$$

Based on these statistics and similar measures associated with uniform variables, we can get the threshold and constant vector estimator for the frequency estimation results

$$g_\theta(\mathbf{X}) = \left[ \mathbb{E}[w_0], ..., \ \mathbb{E}[w_0] + \frac{m-1}{N} + \mathbb{E}[w_m] \right] \quad (41)$$

$$= \left[ 0.125, \ ..., \ 0.125 + \frac{(m-1)}{N} + \sqrt{\frac{5}{N\pi}} \right]. \quad (42)$$

The independence of $w_i$ is used in (41) to separate the expectations. Because $g_\theta(\mathbf{X})$ is an unbiased estimator, the threshold only depends on the variance of the random vector $f_i$. We use $\sigma^2()$ to refer to the variance of a random variable, so the threshold is

$$L_{\text{threshold},\mathbf{f}}(m) = \frac{1}{m}\sum_{i=1}^{m} \sigma^2(f_i) \quad (43)$$

$$= \frac{1}{m}\sum_{i=1}^{m}\frac{1}{64} + \frac{1}{m}\sum_{i=2}^{m}\frac{5}{2N}\left(1 - \frac{2}{\pi}\right) \quad (44)$$

$$= \frac{1}{64} + \left(1 - \frac{1}{m}\right)\frac{5}{2N}\left(1 - \frac{2}{\pi}\right). \quad (45)$$

Here we average over all samples and outputs for multiple output mean squared error to produce a scalar loss value. This results in $L_{\text{threshold},\mathbf{f}} \approx [-18\text{dB}, -16.4\text{dB}, -16\text{dB}, -15.8\text{dB}, -15.7\text{dB}]$ for each value of $1 \le m \le M$. While there is potential for the max-cutoff at $f = 0.5$ to cause the distributions to be heavy-tailed (see e.g. Fig. 5), this does not occur in expectation, so we do not include it in the simplified learning threshold. It can be seen, however, that it does have an effect on the distributions for $m = \{4, 5\}$ in Figure 5.

Our results show that, especially for $m \in \{2, 3\}$, our algorithm consistently outperforms the Periodogram by $3 - 8\text{dB}$, and is only surpassed for the $m = 1$ case with SNR $> -4\text{dB}$. Given more data, it is likely that our algorithm would reach or surpass the Periodogram consistently, however we only train on a small dataset to focus on the insights and learning of our architecture. Our results are especially interesting, because the potential gain of using our algorithm for two- or three-sinusoid signals is an order of magnitude better performance.

### D. Amplitude estimation

Next, we consider the amplitude estimates of our algorithm. While the frequency estimates are resolvable for any bit resolution with sufficiently many samples $N$ and reasonable spacing, the amplitude estimates can be particularly challenging near critical frequencies [28]. For example, consider the following: a single sinusoid with normalized frequency of $f = 0.25 \pm \epsilon$, where $\epsilon < 1/N$, $\phi = 0.1$, and one-bit quantization. In this setting, every subset of $1/f = 4$ points are identical and only contain $\{1 + j, -1 + j, -1 - j, 1 - j\}$. Thus the received sequence, without noise, will be $N/4$ repetitions of that sequence, which makes accurate amplitudes estimation infeasible. While this is true for frequency estimation as well, the repetition improves the frequency estimate, and as $N \to \infty$ the frequency estimate will converge to $f$. There is no similar guarantee for the amplitude estimates, because regardless of the amplitude, the observation sequence is still the same. Even with infinite SNR very little amplitude information is recoverable from such limited data. Nevertheless, the learning threshold does not depend on the input data, so we can still analyze the expected worst case non-adversarial results. Because $a_i$ are independent, the threshold is simply

$$\mathbb{E}[a_i] = 0.55$$

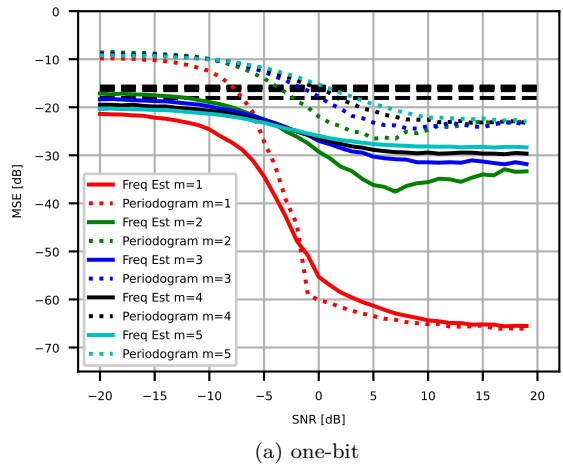

(a) one-bit

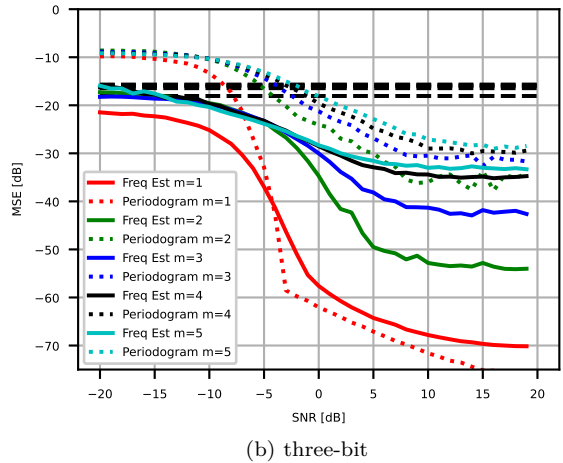

(b) three-bit

Fig. 9. Two plots of the results of our sinusoidal estimation module's frequency outputs (Freq Mod –) for a given m with one-and three-bit data. Our estimator outperforms the threshold universally, and the periodogram (. . .) except in the $m = 1$ case for SNR $\geq -4$dB. For $m > 1$, our estimator is consistently better than the periodogram. In (a) both our networks and the Periodogram experience worse performance for high SNR, which we also observed in our past work with one-bit quantization [12]. Our results only show the performance loss for $m = 2$, while the Periodogram converges to $\approx -23$dB for all $m > 1$. The results in (b) no longer show regressing performance and the increased data resolution especially improves the $m = 2, 3$ cases, where our estimator improves upon the Periodogram by 11-15dB.

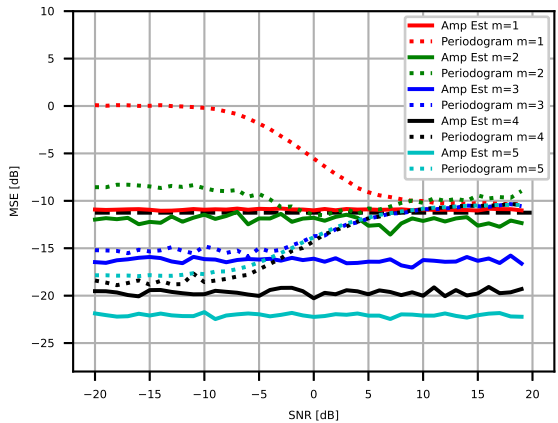

(a)

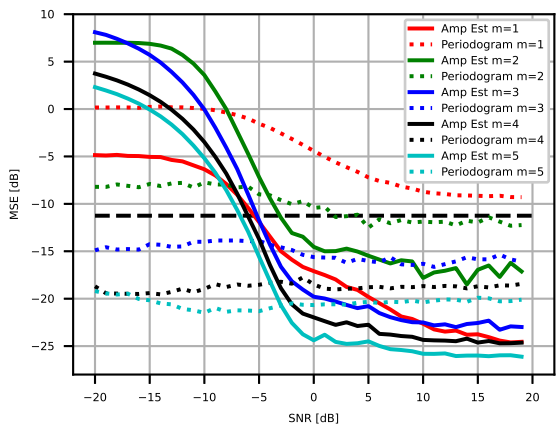

(b)

Fig. 10. A comparison of our sinusoid estimator's amplitude results (Amp Est −). Unexpectedly, our estimator appears to perform better for one-bit data than three-bit at low SNR. Identifying the learning threshold, however, shows the Periodogram and our m = {1, 2} estimators converging to a similar level in (a). In contrast, all of our estimators surpass the threshold for three-bit data in (b).

$$L_{\text{threshold},\mathbf{a}}(\text{m}) = \frac{1}{\text{m}} \sum_{i=1}^{m} \sigma^2(a_i)$$
$$= \sigma^2(a_1) = 0.0675.$$

While the amplitude learning threshold is on a similar scale to the frequency learning threshold, actually achieving this value is challenging for low resolution due to quantization effects directly impacting the amplitude of the signal. We will again rely on the Periodogram as a benchmark for amplitude estimation, although it is not known whether other algorithms have been shown to be more effective for quantized sinusoids.

From Figure 10, we obtain insight into what is happening between the one- and three-bit versions. In 10a most of the estimators are not learning any meaningful features from the input and are simply minimizing the loss over the distribution, based on the lack of improvement with increasing SNR. In 10b, the estimators perform worse initially because of the quantization and noise effects, but outperform the threshold for SNR > −3dB. Thus our three-bit results are much more useful and generalizable than the one-bit case, even if the performance is initially worse for low SNR. From these plots in particular, we can see the value of defining the learning threshold when evaluating deep learning algorithms. Interestingly, increasing the number of sinusoids to m = {3, 4, 5} leads to better results, even though the true amplitudes are independent. We believe this is because as m increases, the average error tends toward the arithmetic mean of each the amplitude estimates, but in the case of m = {2, 3}, if the minimum separation is not sufficient, the estimation error is dominated by the overlapping components.

### E. Phase estimation

In the final evaluation of the sinusoid estimator, we look at the phase estimates. We begin in a similar fashion to the amplitude and phase results by first solving for the learning threshold. Similar to the amplitude distribution, the phases are uniformly distributed, $\phi_i \in [0, 2\pi]$, so the constant estimator and learning threshold are simply

$$\mathbb{E}[\phi_i] = \pi$$
$$L_{\text{threshold},\boldsymbol{\phi}}(m) = \frac{1}{m} \sum_{i=1}^{m} \sigma^2(\phi_i)$$
$$= \sigma^2(\phi_1) = \frac{\pi^2}{3}.$$

From the learning thresholds, we expect the phase error to be significantly higher than the other estimator losses, which is why we scale the sum training loss according to the learning thresholds in Section IV. This should help ensure none of the losses affect the model significantly more than the others, based on the simulation results.

As mentioned in Section V, the Periodogram is replaced with the FFT as a benchmark for estimating the phase of a signal. We show the results of this algorithm with ours in Figure 11, this time without dB scaling. We can see that the one-bit results in Fig. 11a are better than the amplitude estimates, in the sense that every estimator is able to learn and improve with increasing SNR. Similar to the results from Figure 9b, only the m = 1 case shows the benchmark (FFT) performing better than our estimator. In general, the FFT does not perform well because of the short sample lengths $N = 64$, rather than the quantization. In the three-bit results, Fig. 11b, we see that the FFT results are largely unchanged from the one-bit version, but our estimator now shows a noticeable improvement with increasing SNR. We also show the case where we increase $N \to 1024$ for the two-sinusoid FFT to show that sample length is the limiting factor.

### F. SignalNet results

In the final results, we join the two modules, and consider pairings with different traditional estimators. We no longer show the learning threshold, instead comparing across different combinations of algorithms to see how effective our algorithm is in the overall performance. We start by demonstrating the effects of normalization on SignalNet. As we previously found, the phase estimates tend to have much higher error, simply because the phase values are drawn from a distribution with a higher variance. Because the Chamfer loss, assuming the correct number of sinusoids ($m = \hat{m}$), is just two times the mean absolute error, we can approximate the thresholds by the root mean squared error, which is an upper bound on the mean absolute error. Thus, we normalize by the square root of the learning thresholds from Section VIII.

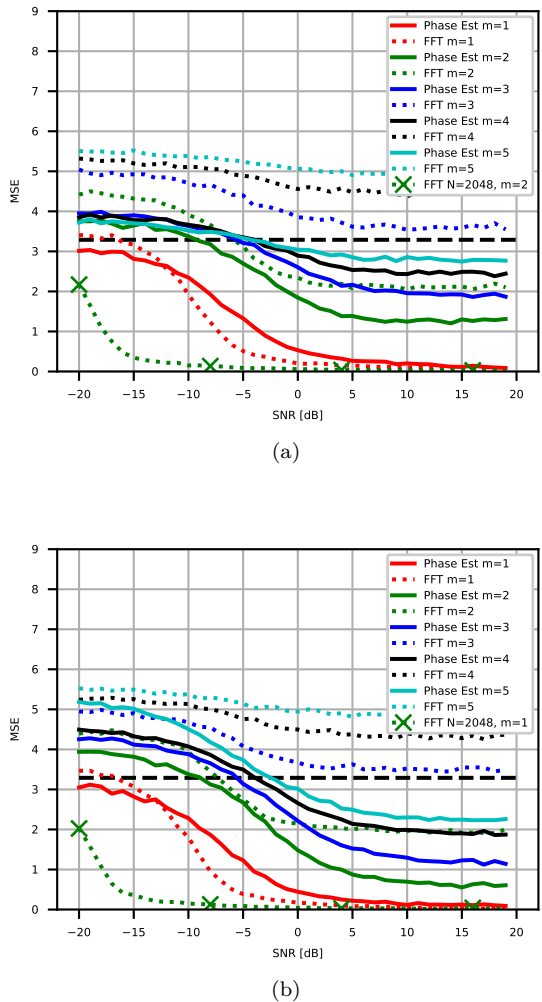

(a)

(b)

Fig. 11. A comparison of the estimation performance of our estimator and the FFT for (a) one-bit data and (b) three bit data. The chart on the left shows all of the estimators achieving and passing the threshold beyond SNR = −5dB, however most of the estimators have only a slight improvement over the threshold. In contrast, our estimator in (b) is able to successfully surpass the threshold for SNR > −3dB for every value of m. Interestingly, the FFT does not noticeably improve with the data resolution, however we found that this was due to the length $N = 64$ causing the phase estimates to be too coarse for the input data. We can see that in (b) the $N = 1024$ phase estimate is slightly improved from (a), and significantly improved from the $N = 64$ setting.

After normalizing, we evaluate benchmarks using AIC

with the Periodogram/FFT against SignalNet. AIC is chosen because of its success over MDL in Figure 6. We interchange AIC with our detection module as well and interchange the Periodogram/FFT with our sinusoidal estimator to identify which components provide the most gain in Figure 12. We color in the two regions corresponding to the gain from using our sinusoidal estimator (red \\ shading) and our detection module (blue // shading).

The final results in Fig. 12, show the sinusoid estimation module providing the most noticeable gain. We can also see that the combination is not uniform: switching to our detection module from AIC provides significant gain, but so does switching to our sinusoid estimator from the Periodogram/FFT. This is because our sinusoid estimator has learned to fit the best estimates for a given number of sinusoids, so even with poor detection results, it still has reasonable Chamfer loss. This explains the asymmetric gain from adding or subtracting one of our modules between the AIC+FFT curve and our SignalNet curve. Recall however, that the learning threshold normalization has been applied to the estimation module, but not to the detection module, as there is no intuitive way to apply that with the Chamfer loss definition. Based on the results from the one-bit detection module in Fig. 6, a significant portion of the gain being seen by the detection module can be attributed to learning the distribution, rather than learning that generalizes beyond our simulation. Ultimately, our SignalNet architecture appears to provide a valuable improvement over other techniques, when compared in our simulation setup. However, we cannot give conclusive statements until we also evaluate how our algorithm generalizes to different data distributions.

### G. Generalizability

In our final results, we evaluate the out-of-distribution (OOD) results of our algorithm, determined from uniformly distributed frequency content. Given the structure of our generation process in Algorithm 1, we wish to also have a measure of the performance on more general data settings. As we mentioned in Section IV, it is crucial that we generate frequency content that is separated by at least $1/N$, but also should not be equally spaced. In this section, we generate the sinusoidal frequencies from a uniform distribution, with the restriction that no frequency content can be within $1/N$ of any other. If there is insufficient spacing, we regenerate an entirely new set. For each value of $m$, we draw frequencies from a uniform distribution with the same domain as our original estimator, to ensure that the learned domain remains consistent.

We show the results of all three estimators for the $m = 2$ case, which is the distribution furthest from a uniform distribution, and therefore experiences the greatest distribution shift. We can determine which case has the greatest distribution shift through a variety of metrics, such as a discrete Kolmogorov-Smirnov test, but it is most intuitive to compare the distributions in Fig. 5 with a uniform distribution over the same domain. We can see from the results in Figure 13, that there is almost no performance impact due to the change in distribution. Based on the ability of our neural network to generalize and outperform the benchmarks, we can conclude that applying the SignalNet architecture to the detection and estimation task is a valuable improvement over other techniques.

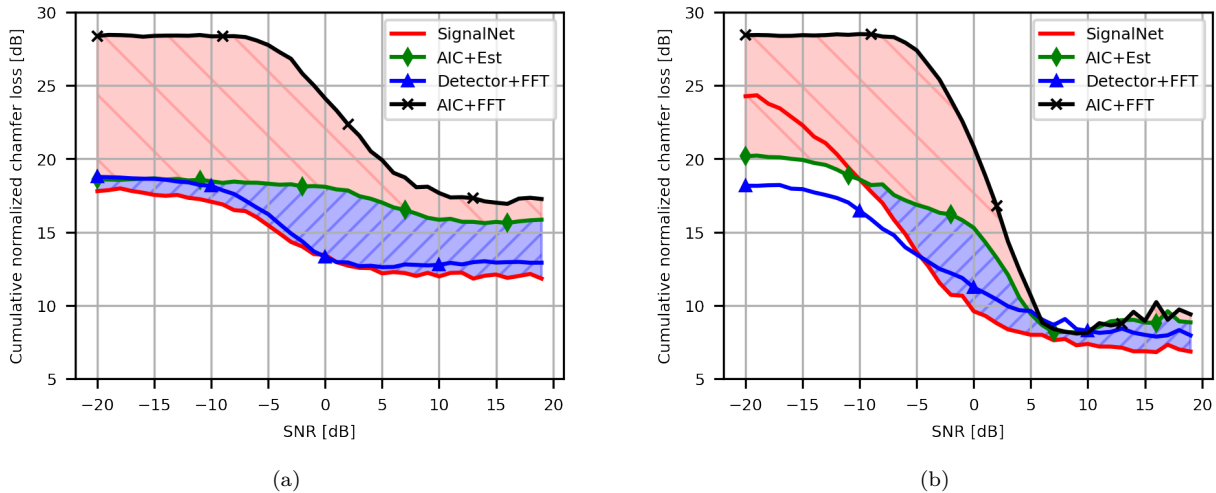

Fig. 12. The resulting chamfer loss for each of the algorithms with one-bit data (a) and three-bit data (b). We color two regions: one region to show the gain from our detector (blue //) and the other to show the gain from our sinusoid estimator (red \\). At low SNR, the estimator provides significant gains, in part from distributional learning, while at high SNR, the detector provides a larger gain due to the limitations of low resolution sampling for AIC (Fig. 6).

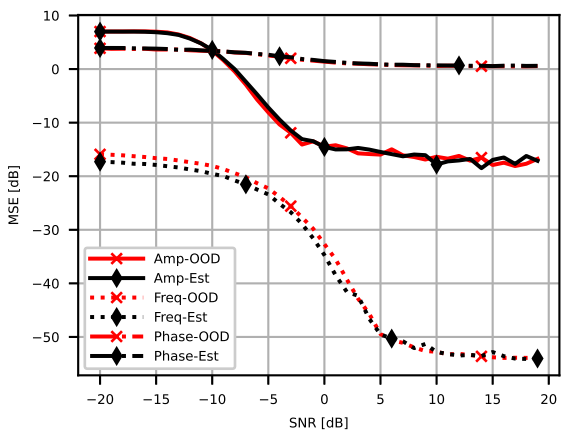

Fig. 13. A comparison of our proposed network testing with the training distribution (black) and out-of-distribution samples (red) for the m = 2 case. Frequency content is drawn from a uniform distribution with the same domain as the training data. The results are almost identical, showing that the features learned by our model are effective, even under distribution shift.

## IX. Conclusion

In this paper, we described and evaluated SignalNet, a novel deep neural network for multi-sinusoidal decomposition from low resolution sampling. While no other work has considered quantized, multi-sinusoidal decomposition, our network follows traditional work by dividing the problem into the subtasks of detection and estimation. We describe a distinct architecture for each subtask and specifically focus on developing a novel estimation network. Our estimation network uses internal reconstruction to explicitly learn the input-output relationship. Our algorithm struggled to learn from one-bit data in both detection and amplitude estimation, though, similar to the baseline algorithms. Yet, we also saw the benefit that instilling domain knowledge in the form of reconstructive information provided to our estimator, allowing it to efficiently learn features from the data in all other tests. While our results strictly consider the case of estimating all sinusoidal parameters, in the case that a specific parameter is not necessary or is known a priori, a constant vector can be used in its place during reconstruction.

We also proposed a theoretical tool for comparing our networks and normalizing across distributions by defining a learning threshold. The threshold is used to express the randomness in the output variable for a specific loss function. We used the insight from the learning threshold to understand why the performance of our model did not universally increase going from the one-bit to the three-bit versions and often had better low SNR performance with one-bit data–by not learning meaningful features. This prevented the estimator from performing well, but also made it robust to noise. We found that our learning threshold is a useful metric to consider when comparing neural networks with broader algorithms and determining the success and generalizability of learning algorithms.

Finally, we combined our networks together to complete the SignalNet architecture. Our unified network is able to surpass the benchmark algorithms universally. We were able to directly quantify the improvement from our sinusoid estimator which provided between 2-10dB improvement. We conclude that a domain-aware detection module could improve the results further, as well as additional investigation into appropriate loss functions. Our results suggest that multi-sinusoid decomposition can be performed even with extremely low resolution quantization using our SignalNet architecture.